\documentclass[draft,a4paper,prb,amssymb,showpacs]{revtex4}
\input epsf

\begin{document}

\title{Does theory of quantum correction to conductivity agree
 with experimental data in 2D systems?}
\author{G.~M.~Minkov}
\email{Grigori.Minkov@usu.ru}
\author{O.~E.~Rut}
\author{A.~V.~Germanenko}
\author{A.~A.~Sherstobitov}
\affiliation{Institute of Physics and Applied Mathematics, Ural
State University, 620083 Ekaterinburg, Russia}
\author{V.~I.~Shashkin}
\author{O.~I.~Khrykin}
\author{V.~M.~Daniltsev}
\affiliation{Institute of Physics of Microstructures of RSA,\\
603600 Nizhni Novgorod, Russia}

\date{\today}

\begin{abstract}
The quantum corrections to the conductivity have been studied in
the two types of 2D heterostructures: with doped quantum well and
doped barriers. The consistent analysis shows that in the
structures where electrons occupy the states in quantum well only,
all the temperature and magnetic field dependences of the
components of resistivity tensor are well described by the
theories of quantum corrections. Contribution of the
electron-electron interaction to the conductivity has been
determined reliably for the structures with different electron
density. A possible reason of large scatter in experimental data
relating to the contribution of electron-electron interaction,
obtained in previous papers, and the role of the carriers,
occupied the states of the doped layers, is discussed.
\end{abstract}

\maketitle

\section{Introduction}

\label{sec1}

The quantum corrections to the Drude conductivity in disordered
metals and doped  semiconductors are intensively studied since
1980.\cite{r0,r01} Two mechanisms lead to these corrections: (i)
the interference of the electron waves propagating in opposite
directions along closed paths; (ii) the electron-electron
interaction (EEI).  These corrections increase with decreasing
temperature and/or increasing disorder, and the low temperature
transport in 2D systems is largely determined by those. Recently,
two different behaviors of the conductivity of 2D system with
decrease of temperature $T$ have come to light: (i) the
conductivity decreases monotonically for one type of systems; (ii)
the conductivity decreases at sufficiently high temperature, but
reveals surprising growth  at low enough temperature for other
ones. \cite{r1}

It is commonly accepted that decrease of the conductivity for the
first type of systems results from temperature dependence of the
quantum corrections that are negative and logarithmically diverge
at $T\rightarrow 0$. In such systems the crossover from the weak
localization (WL) regime, when the corrections are small compared
with the Drude conductivity, to the strong localization (SL)
regime is observed. The role of interference and electron-electron
interaction in this crossover is of special interest and attracts
much attention in recent years.

As for the second type of the 2D systems, no general consensus on
the origin of metallic behavior has been reached as yet. The study
of the role of the EEI and interference can be useful for
understanding the origin of the metallic-like behavior of
conductivity in such systems.

\begin{figure}[b]
\epsfclipon
 \epsfxsize=8cm
 \epsfbox{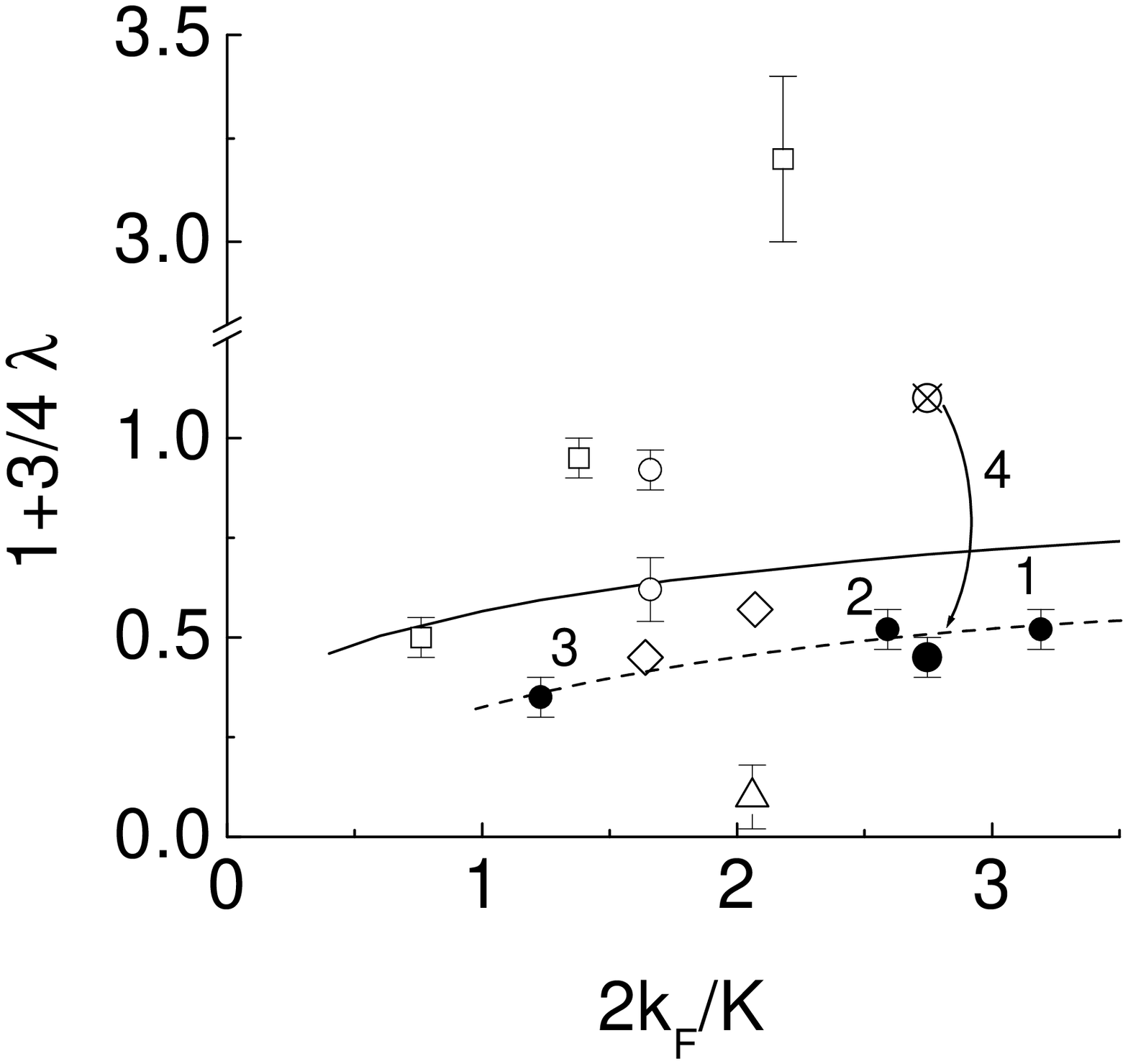}
\caption{ The value of multiplier $\left( 1+3/4\protect\lambda
\right) $ in Eq.~(\protect\ref{eq01}) as function of $k_{F}/K$.
Symbols are the experimental results from
Refs.~\protect\onlinecite{Tsui} ($\square$),
\protect\onlinecite{Polyanskaya} ($ \bigcirc $),
\protect\onlinecite{Poirier}($\triangle $),
\protect\onlinecite{r6}($\lozenge$) , and our data ($\bullet $,
$\otimes $). The solid curve represents the theoretical result
from Ref.~\protect\onlinecite{r5}, dotted line is the guide for an
eye. Arrow indicates the shift of experimental point for structure
4 after extraction of temperature dependence of electron mobility
(see Section \protect\ref{sec:disc}).} \label{fig1}
\end{figure}

The WL-SL crossover was intensively studied in thin metal films.
It is generally assumed that the EEI has a crucial role because
the interference is suppressed by the strong spin-orbit
interaction in metals. Different aspects of the WL-SL crossover
were studied in semiconductor 2D structures but there is no
conventional view on the role of EEI and interference in this
crossover up to now. Moreover, the magnitudes of the EEI and
interference corrections to the conductivity are not well
established experimentally in the WL regime, when theories of the
quantum corrections are applicable. It is especially concerns the
EEI contribution. It was shown in Ref.~\onlinecite{r0,rEEI} that
the EEI contribute to $\sigma_{xx}$ only. For $g\mu_B B/kT\lesssim
1$ the correction has the form
\begin{equation}
\Delta\sigma_{xx}^{ee}=G_0
\left(1+\frac{3}{4}\lambda\right)\ln\left(\frac{
kT\tau}{\hbar}\right),  \label{eq0}
\end{equation}
whereas for $g\mu_B B/kT\gg 1$ it is given by
\begin{equation}
\Delta\sigma_{xx}^{ee}=G_0
\left(1+\frac{1}{4}\lambda\right)\ln\left(\frac{
kT\tau}{\hbar}\right).  \label{eq01}
\end{equation}
Here, $ G_0=e^2/(2\pi^2\hbar)$, $\tau$ is the momentum relaxation
time, $\lambda$ is a function of $k_F/K$\cite{r5} with $k_F$ as
the Fermi quasimomentum and $ K $ as the screening parameter,
which for 2D case is equal to $2/a_B$, where $a_B$ is the
effective Bohr radius. For the most intensively studied Si
MOS-structures, the transition to the case $g\mu_B B/kT\gg 1$
occurs within the actual range of temperature and magnetic field.
It results in complicated magnetic field dependence of the
resistance: decreasing at low magnetic field it increases at high
ones. This makes the quantitative interpretation of experimental
data difficult.

For electron 2D systems based on GaAs, in which the electron
$g$-factor is much smaller than that in Si, the condition $g\mu_B
B/kT \lesssim 1$ is fulfilled in wide range of temperature and
magnetic field except extremely low temperature or high magnetic
field. Therefore, the experimental results can be interpreted in
the most simple way for these systems. The multiplier before
logarithm in Eq.~(\ref{eq0}) is determined experimentally and just
its value is shown in Fig.~\ref{fig1} as function of $k_F/K$.
Theoretical curve from Ref.~\onlinecite{r5} is also shown in the
figure. The large scatter of the experimental data shows that
there are no reliable data on the contribution of the EEI and it
is impossible to conclude whether the theory describes the
experiment.

From our point of view, before the discussion of the very
interesting problem concerning the role of the EEI in WL-SL
crossover and in the new low temperature metallic phase, it would
be essential to acquire reliable data in the WL regime far from
the WL-SL crossover. Just this problem our paper is devoted to.

Let us discuss what type of semiconductor heterostructures would
be the most suitable for quantitative study of the quantum
corrections to the conductivity in the 2D systems. First of all,
the Drude conductivity $\sigma_0$ should be high: $\sigma_0/(\pi
G_0)=k_F l\gg 1$, where $l$ is mean-free path. In this case the
WL-theory can be applied.  On the other hand, the quantum
corrections must not be very small, lest the measurement accuracy
restricts the quantitative analysis. This means that the electron
scattering  must be strong enough, i.e., the mobility must not be
high. It should be the single-quantum-well heterostructure with a
single subband occupied, as the theories of quantum corrections
have been developed mainly for such case. Quantum well should be
symmetric in shape. It allows to eliminate the peculiarities
caused by the spin-orbit interaction\cite{hik,spin} and to neglect
the spin effects under analysis of experimental data. It should be
the structure with electrons in quantum well only, i.e. with empty
doped (modulation- or $\delta$-) layer. It enables one to avoid
the shunting of the quantum well. The role of the carriers in
doped layers is not restricted by shunting. Their redistribution
within the layers with temperature can lead to the
temperature-dependent disorder, and, hence, to additional
temperature dependence of the mobility. \cite{TDdisorder}

Thus, two types of structures meet these requirements: (i) the
structures with doped quantum well; (ii) the structures with
symmetrically doped barriers and low carrier density, when the
Fermi level lies significantly lower than any states in doped
layers. Exactly these types of structures was studied in this
work.

\section{Theoretical basis}

\label{sec2}

In this section we present the main theoretical results which will
be used in the analysis of the experimental data.\cite{r0} These
theories are valid when $k_F l\gg 1$ and the quantum corrections
to the conductivity are small compared with the Drude
conductivity.

Without an external magnetic field the total quantum correction to
conductivity is
\begin{equation}
\delta\sigma(T)=G_0\left[\ln\left(\frac{\tau}{\tau_\varphi(T)}\right)
+\left(1+\frac{3}{4}\lambda\right)\ln\left(\frac{kT\tau}{\hbar}\right)\right],
\label{eq011}
\end{equation}
where $\tau_\varphi$ is the phase-breaking time. The first term in
Eq.~(\ref {eq011}) is the interference correction, the second one
is the EEI contribution. At low temperatures the phase-breaking
time is determined by inelasticity of the electron-electron
interaction and is
\begin{equation}
\tau_\varphi^{-1}= \frac{kT}{\hbar}\frac{2\pi G_0}{\sigma_0}\ln\left(\frac{%
\sigma_0}{2\pi G_0}\right).  \label{eq08}
\end{equation}
The value of $\lambda$ was obtained in Ref.~\onlinecite{r5}
\begin{equation}
\lambda=4\left[1-2\frac{\left(1+\frac{1}{2}F\right)\ln {\left(1+\frac{1}{2}%
F\right)}}{F}\right],  \label{eq03}
\end{equation}
where
\begin{equation}
F=\int{\frac{d \theta}{2\pi}\left[1+\frac{2k_F}{K}\sin{\frac{\theta}{2}}%
\right]^{-1}}.  \label{eq031}
\end{equation}

In a magnetic field the classical conductivity tensor has the
following form:
\begin{eqnarray}
\sigma_{xx}^0=\frac{en\mu}{1+\mu^2 B^2},  \label{eq021} \\
\sigma_{xy}^0=\frac{en\mu^2 B}{1+\mu^2 B^2}.  \label{eq02}
\end{eqnarray}
The electron-electron interaction contributes to $\sigma_{xx}$
only [see Eqs.~(\ref{eq0}) and (\ref{eq01}) for
$\Delta\sigma_{xx}^{ee}$], whereas $\Delta\sigma_{xy}^{ee}=0$. It
is easy to show that the magnetoresistance
\begin{equation}
\rho_{xx}(B,T)=\frac{\sigma_{xx}^0(B)+ \Delta\sigma_{xx}^{ee}(T)}{%
(\sigma_{xy}^0(B))^2+(\sigma_{xx}^0(B)+ \Delta\sigma_{xx}^{ee}(T))^2}
\label{eq04}
\end{equation}
is parabolic in the form when $\Delta\sigma_{xx}^{ee}\ll \sigma_{xx}^0$:
\begin{equation}
\rho_{xx}(B,T)\simeq \frac{1}{\sigma_0}-\frac{1}{\sigma_0^2}\left(1-\mu^2
B^2\right)\Delta\sigma_{xx}^{ee}(T).  \label{eq05}
\end{equation}
So, $\rho_{xx}$-versus-$B$ curves for different temperatures should cross
one another at fixed point $B_{cr}=1/\mu$ and the value of $%
\rho_{xx}^{-1}(B_{cr})$ should be equal to the Drude conductivity.\cite
{DyakCondMat}

\begin{figure}[t]
\epsfclipon
 \epsfxsize=8cm
 \epsfbox{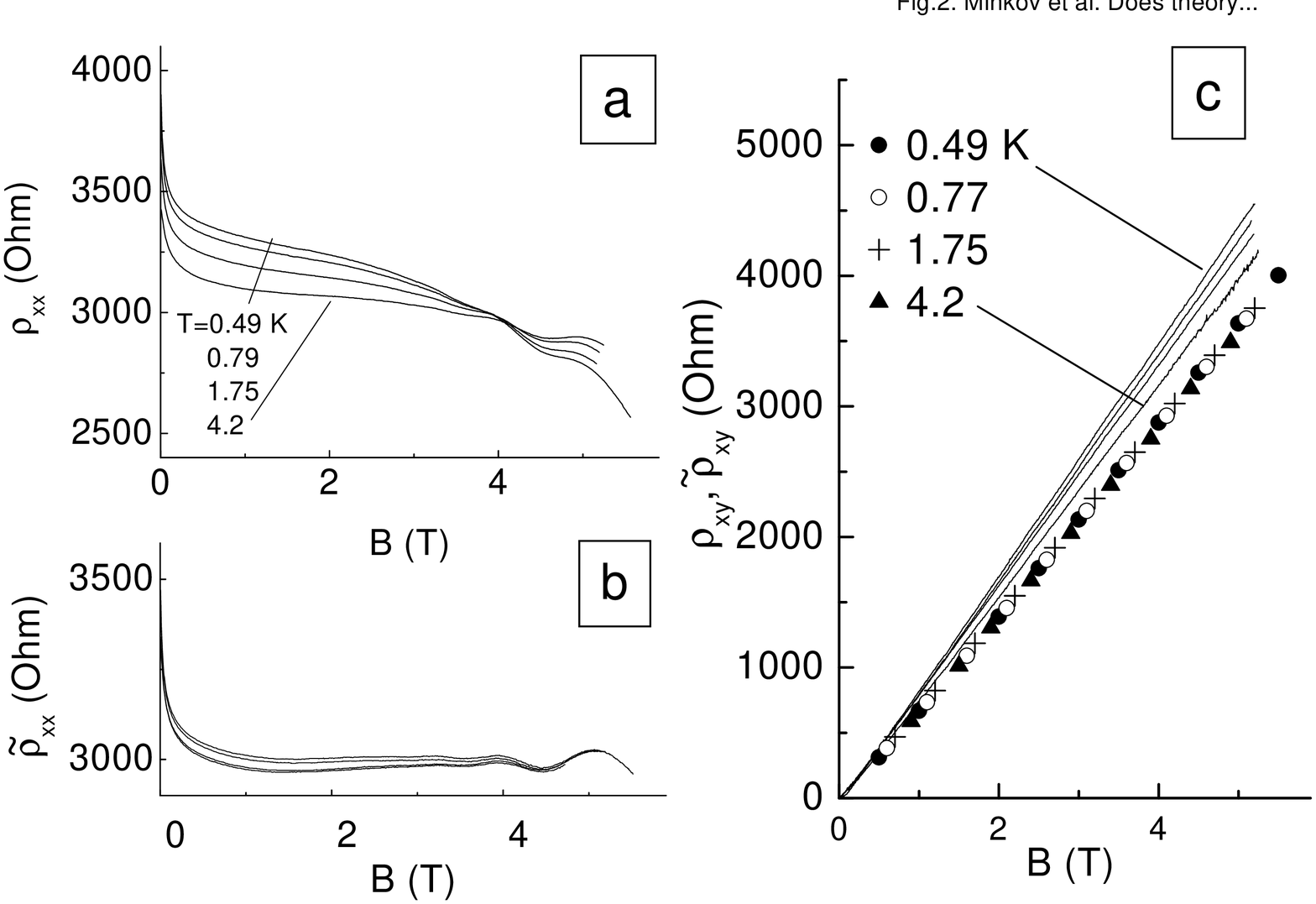}
\caption{Magnetic field dependences of $\protect\rho _{xx}$ (a), $\tilde{%
\protect\rho}_{xx}$(b) and $\protect\rho _{xy}$ (lines), $\tilde{\protect\rho%
}_{xy}$ (symbols) (c) for different temperatures for structure 2.}
\label{fig2}
\end{figure}

The interference correction to the conductivity gives the
contributions both to $\sigma_{xx}$ and $\sigma_{xy}$ but their
ratio is such that $\rho_{xy}$ remains unchanged. Within the
framework of the diffusion approximation, which is valid when
$\tau_\varphi/\tau\gg 1$ and $B <B_{tr}=\hbar c/(2el^2)$, the
magnetic field dependence of $\Delta(1/\rho_{xx}^{in})=1/\rho_{xx}(B)-1/%
\rho(0)$ is described by the well-known expression\cite{hik}
\begin{equation}
\Delta(1/\rho^{int}_{xx}(B))=\alpha G_0 \biggl\{ \psi\left(\frac{1}{2}+\frac{%
\tau}{\tau_\varphi}\frac{B_{tr}}{B}\right) - \psi\left(\frac{1}{2}+\frac{%
B_{tr}}{B}\right)- \ln{\left(\frac{\tau}{\tau_\varphi}\right)} \biggr\},
\label{eq07}
\end{equation}
where $\psi(x)$ is a digamma function, the value of $\alpha$ is
equal to unity. The magnetic field dependence of the interference
correction beyond the diffusion approximation was studied in
Refs.~ \onlinecite{chak,schm,dyak,dmit,novok,our1}. The analytical
expression suitable for fitting of the experimental data was not
obtained for this case, however, as is shown in
Ref.~\onlinecite{our1}, Eq.~(\ref{eq07}) also describes the
$\Delta(1/\rho^{in}_{xx})$-vs-$B$ curve well  but with prefactor
$\alpha<1$.

It follows from Eqs.~(\ref{eq05}) and (\ref{eq07}) that the interference
correction gives the strong magnetic field dependence of the resistivity at $%
B\leq B_{tr}$, whereas the EEI does it at magnetic field $B\geq B_{cr}=1/\mu$%
. Since the ratio $B_{cr}/B_{tr}$ is equal to $2k_F l$, these
magnetic field ranges are well separated. Thus, application of
magnetic field allows to obtain the interference and interaction
contributions to the conductivity separately.

To answer the question ``Does the theory of quantum corrections
agree with experiment in 2D systems?", all the theoretical
predictions given above should be checked step by step.
\begin{figure}
\epsfclipon
 \epsfxsize=8cm
 \epsfbox{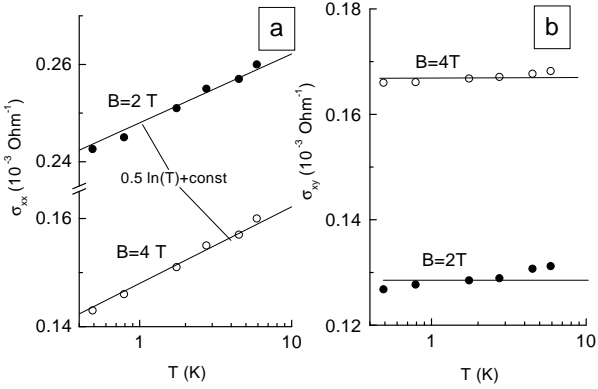}
\caption{Temperature dependence of $\protect\sigma_{xx}$ (a) and $\protect%
\sigma_{xy}$ (b) for two magnetic fields for structure 2.}
\label{fig3}
\end{figure}

\begin{table*}[tbp]
\caption{Sample parameters}
\begin{tabular}{ccccccccc}
Str. & $\sigma _{0}$($10^{-4}$ Ohm$^{-1}$) & n($10^{12}$ cm$^{-2})$ & $\tau $%
($10^{-14}$sec) & $B_{tr}$(T) & $k_{F}l$ & $B_{cr}$ (T) & $\rho
_{xx}^{-1}(B_{cr})$(Ohm$^{-1}$) & $\mu ^{-1}$\footnote{The value
of
mobility has been determined as $\mu=\rho_{xy}/(\rho_{xx}B)$ at $B=B_{cr}$.}%
(T) \\
\colrule 1 & $(4.13\pm 0.02)$ & $1.35\pm 0.05$ & $6.5$ & 0.25 &
10.7 & 5.22
& $3.95\times 10^{-4}$ & 5.26 \\
2 & $(3.55\pm 0.03)$ & $0.87\pm 0.02$ & $8.8$ & 0.21 & 9.2 & 4.15 & $%
3.38\times 10^{-4}$ & 4.17 \\
3 & $(1.90\pm 0.05)$ & $0.19\pm 0.02$ & $21.0$ & 0.16 & 4.9 & 1.66 & $%
1.67\times 10^{-4}$ & 1.64 \\
4 & $(6.50\pm 0.05)$ & $1.0\pm 0.05$ & $13.7$ & 0.076 & 16.6 & 4.14 & $%
6.45\times 10^{-4}$ & 2.50\\
\botrule
\end{tabular}
\label{tab1}
\end{table*}

\section{Samples}

The heterostructures with 50\AA\ In$_{0.15}$Ga$_{0.85}$As single quantum
well in GaAs with Si $\delta$-doping layers were investigated. Two types of
heterostructures were studied: with doped quantum well (structures 1 and 2),
and with doped barrier (structures 3 and 4). In the first case $\delta$%
-layer was arranged in the center of quantum well. In the second one, two $%
\delta$-layers, separated by the 60\AA\ GaAs spacer, were disposed on each
side of the quantum well. The thickness of undoped GaAs cap layer was 3000
\AA\ for all structures. The samples were mesa etched into standard Hall
bridges. The parameters of the structures are presented in Table~\ref{tab1}.

\section{Temperature dependence of conductivity at high magnetic field.
Contribution of electron-electron interaction.}

\label{sec:hf}

The experimental magnetic field dependences of $\rho_{xx}$ and
$\rho_{xy}$ are presented in Fig.~\ref{fig2}(a), (c) for one of
the structures at different temperatures.  The two different
magnetic field ranges are evident:  the range of sharp dependence
of $\rho_{xx}$ at low field $B\leq 0.5-1$ T, and the range of
moderate dependence, which is close to parabolic, at higher field.
All $\rho_{xx}$-vs-$B$ curves cross each other at fixed
magnetic field $B_{cr}=4.15$ T. This value is close to $1/\mu$ (see Table~%
\ref{tab1}). The Hall resistance is practically linear with
magnetic field. However, despite the strong degeneracy of electron
gas ($E_F/(kT)>100$, where $E_F$ is the Fermi energy) the Hall
resistance decreases with increasing temperature.
Low-magnetic-field behavior of $ \rho_{xx}$ is a consequence of
suppression of the interference correction by magnetic field. This
effect will be discussed below.
\begin{figure}[t]
\epsfclipon
 \epsfxsize=8cm
 \epsfbox{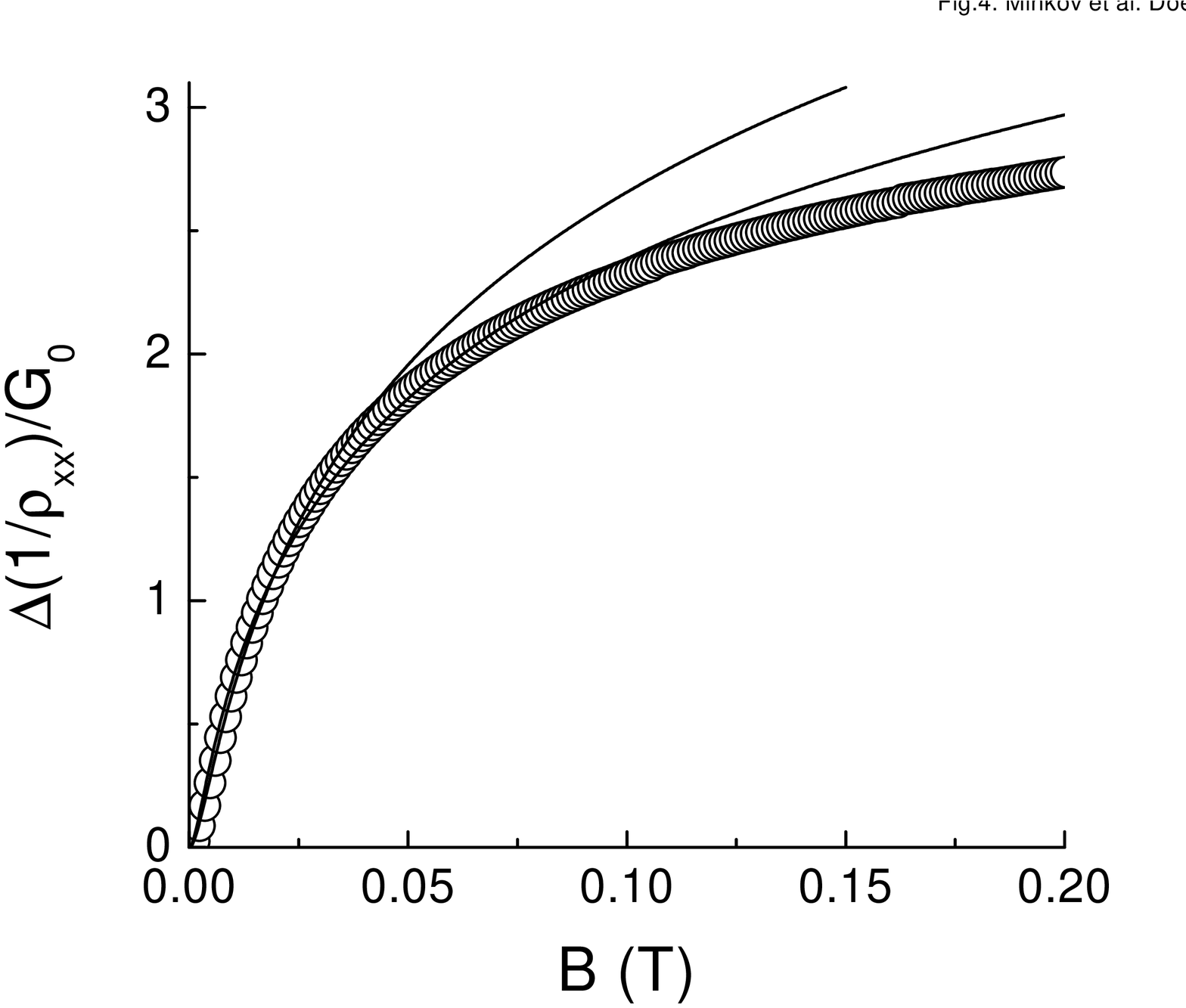}
\caption{The magnetic field dependence of
$\Delta(1/\protect\rho_{xx}(B))$ for structure 2, $T=0.45$ K.
Symbols are the experimental data. Lines are best fit to Eq.
(\ref{eq07}) made over different magnetic field ranges: $\Delta
B=(0-0.1)B_{tr}$ (upper line), and $\Delta B=(0-0.3)B_{tr}$ (lower
line). } \label{fig31}
\end{figure}

At high magnetic field ($B>1-2$ T), $\rho_{xx}(B,T)$ and
$\rho_{xy}(B,T)$
differ from the classical behavior following from Eqs.~(\ref{eq021}) and (%
\ref{eq02}), by contribution of electron-electron interaction
only. To assure in this fact we plot the temperature dependences
of the conductivity tensor components  in Fig.~\ref{fig3}. It is
clearly seen that the change of $\sigma_{xx}$ with temperature
does not depend on $B$ and significantly larger than that of
$\sigma_{xy}$. Namely such a behavior is in full agreement with
the predictions of the EEI theory. Thus, the absence of
$\sigma_{xy}$ temperature dependence  allows us to attribute the
$\sigma_{xx}$ temperature dependence to contribution of the EEI
and determine the multiplier before logarithm in Eq.~(\ref{eq0})
from the slope of $\sigma_{xx}$-vs-$\ln{T}$ dependence:
$(1+3/4\lambda)=0.5\pm 0.1 $ (see Fig.~\ref{fig3}(a)). Notice that
practically in all papers where the EEI was studied, $\sigma_{xy}$
temperature independence  was not demonstrated over the magnetic
field range where the EEI contribution was determined. Below we
show that the existence of $\sigma_{xy}$ temperature dependence
introduces a large error into the determination of
$(1+3/4\lambda)$.
\begin{figure}
\epsfclipon
 \epsfxsize=8cm
 \epsfbox{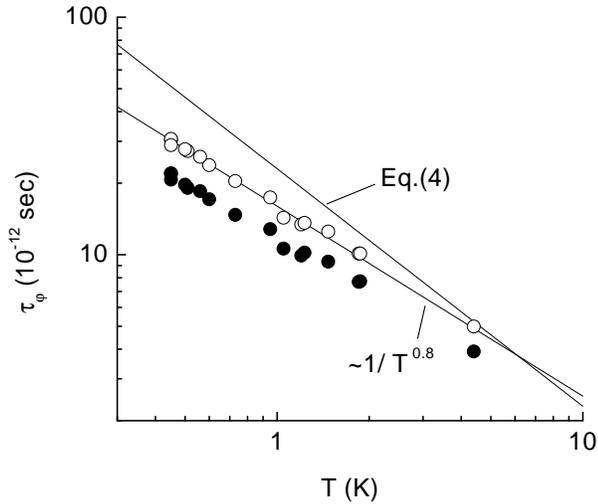}
\caption{Temperature dependence of $\protect\tau_\varphi$.
Symbols are the results of fitting of the experimental data to
Eq. (\ref{eq07}) for different fitting range $\Delta B$: $\Delta
B=(0-0.1)B_{tr}$ (full circles), and $\Delta B=(0-0.3)B_{tr}$
(open circles). } \label{fig4}
\end{figure}

Now when we have determined the EEI contribution let us extract it
from $\sigma_{xx}$, invert the conductivity tensor and plot the components $%
\tilde{\rho}_{xx}$ and $\tilde{\rho}_{xy}$ without this correction (Fig.~\ref
{fig2} (b), (c)). Disappearance of the temperature dependence of $\tilde{\rho%
}_{xy}$ and $\tilde{\rho}_{xx}$ confirms the correctness of
determination of the EEI contribution to the conductivity, and
absence of any mechanisms that can lead to additional temperature
dependence of the conductivity.

Note, that after extraction of the EEI contribution the electron density
determined by the different ways: (i) $B/(e\tilde{\rho}_{xy})$; (ii) $%
B_{cr}/(e\rho_{xx}(B_{cr}))$; and (iii) from the Shubnikov - de Haas
oscillations, are very close to each other and lie within the error interval
given in Table~\ref{tab1}.

\section{The low field magnetoresistance. Interference correction to the
conductivity}

Let us consider the low magnetic field range. The temperature dependence of $%
\rho_{xx}$ in this range is determined by both the EEI and
interference contributions whereas the magnetic field dependence
is determined by interference contribution only, because
$\rho_{xx}(B)$ is unaffected by the EEI at $B\ll 1/\mu$. Thus, the
dependence of $\Delta(1/\rho_{xx}(B))$ must be described by
Eq.~(\ref{eq07}) over this magnetic field range and one can
determine the value of phase-breaking time using $\alpha$ and
$\tau_\varphi$ as fitting parameters. The low-field
magnetoresistance for structure 2 is
presented in Fig.~\ref{fig31}. Detailed analysis of the dependences $%
\Delta(1/\rho_{xx}(B))$ shows that the fitting values of $\alpha$
do not depend on the temperature but to some extent  depend on
the fitting range $\Delta B$: $\alpha=1.2$ at $\Delta
B=(0-0.1)B_{tr}$, and $\alpha=0.9$ at $\Delta
B=(0-0.3)B_{tr}$ (Fig.~\ref{fig4}). The temperature dependences of $%
\tau_\varphi$ are close to $T^{-p}$ with $p\simeq
0.85$\cite{snoska} for both fitting ranges but absolute values of
$\tau_\varphi$ are somewhat different. Theoretical dependences
$\tau_\varphi(T)$ calculated in accordance with Eq.~(\ref{eq08})
are also shown in Fig.~\ref{fig4}. It is seen that magnitudes of
$\tau_\varphi$ are close to theoretical values. Notice that close
to linear $\tau_\varphi$-vs-$1/T$ dependence was observed in most
papers but the magnitude often happened less as much as $3-5$
times. The reasons for that are unclear and one ought to suppose
that additional phase-breaking mechanisms with close temperature
dependence are essential in such structures. Therefore, the
quantitative results for quantum correction, obtained for such
structures seems to be inconclusive.

\begin{figure}
\epsfclipon
 \epsfxsize=8cm
 \epsfbox{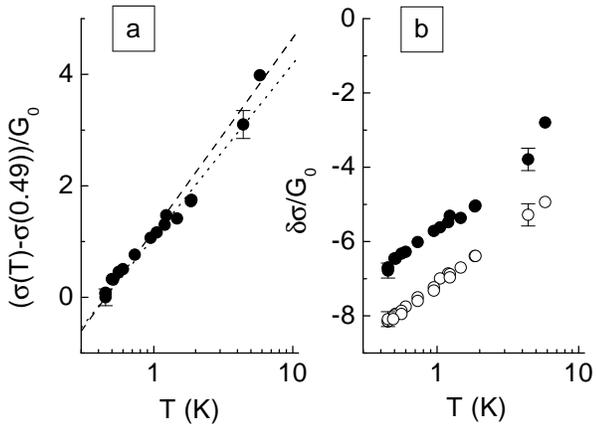}
\caption{(a) The dependence $\Delta \protect\sigma (T)=\protect\sigma (T)-%
\protect\sigma (T_{0})$ with $T_{0}=0.49$~K. Symbols are the
experimental data, lines are given by Eq.~(\ref{eq09}) with
$(1+3/4\protect\lambda )=0.5$ and $p=1$ (dash line), $p=0.85$
(dotted line). (b) The temperature dependence of absolute value of
total quantum correction to the conductivity determined by
different ways (see text).} \label{fig5}
\end{figure}
\begin{figure}[tbp]
\epsfclipon
 \epsfxsize=8cm
 \epsfbox{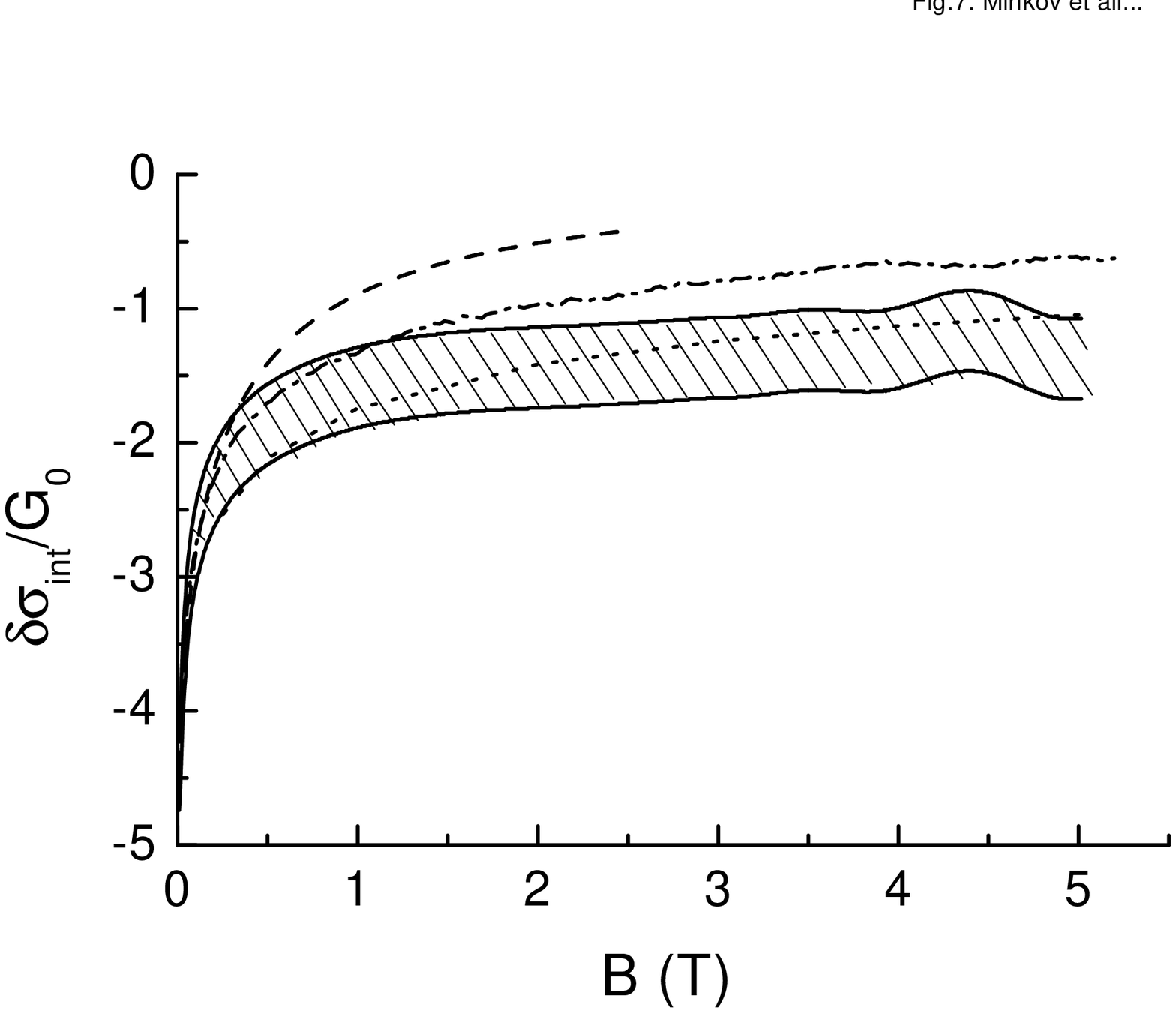}
\caption{The magnetic field dependence of the interference quantum
correction $\protect\delta \protect\sigma _{int}$ over the entire
magnetic field range for structure 2, $T=1.5$~K. The shadowed area
is the experimental result, spread is caused by error in
determination of $\protect\sigma _{0}$ (see Section \ref{sec:b0}).
The dashed line is the result of the diffusion approximation given
by Eq.~(\ref{eq07}), the dot-dashed line takes into account both
the back-scattering and non-back-scattering processes, the dotted
line represents only the back-scattering
contribution.\protect\cite {dmit}} \label{fig6}
\end{figure}

\section{ Temperature dependence of the conductivity at B=0. Absolute
value of the quantum corrections}

\label{sec:b0}

We turn now to temperature dependence of the conductivity at $B=0$ (Fig.~%
\ref{fig5}). It is determined by the temperature dependence of
both the interference correction and correction due to the EEI. As
seen from Eqs.~( \ref{eq011}) and (\ref{eq08}), the relative value
of $\sigma(T)$ has to decrease logarithmically  with decreasing
temperature
\begin{equation}
\Delta\sigma(T)=\sigma(T)-\sigma(T_0) =G_0\left[p\, \ln{\left(\frac{T}{T_0}%
\right)}+\left(1+\frac{3}{4}\lambda\right)\ln{\ \left(\frac{T}{T_0}\right)}%
\right],  \label{eq09}
\end{equation}
where $T_0$ is some arbitrary temperature. Thus, the slope in $\Delta\sigma$%
-vs-$\ln T$ dependence has to be equal to $G_0\left[p+(1+3/4\lambda)\right]$%
. Lines in Fig.~\ref{fig5}(a) show the dependences (\ref{eq09})
calculated with $(1+3/4\lambda)=0.5$ determined above (see Section
\ref{sec:hf}) and
with two values of $p\,$: with theoretical value $p=1$ [see Eq.~(\ref{eq08}%
)], and $p=0.85$ describing the experimental $\tau_\varphi(T)$
dependence (see Fig.~\ref{fig4}). It is evident that the
experimental results coincide with these dependences within
experimental error.

Now let us consider the absolute value of the total quantum
correction $ \delta\sigma$. On the one hand, we can find it from
Eq.~(\ref{eq011}), using the parameters $\tau_\varphi$,
$(1+3/4\lambda)$ determined above, and $ \tau=\mu m/e$, where
$m=0.06\, m_0$ is the electron effective mass in In$
_{0.15}$Ga$_{0.85}$As quantum well. This values are plotted in
Fig.~\ref {fig5}(b) with open circles. On the other hand, the
absolute value of the quantum corrections at given $T$ is equal to
the difference between $ \sigma(T)$ and the Drude conductivity:
$\delta\sigma(T)=\sigma(T)-\sigma_0$. This value obtained with
$1/\rho_{xx}(B_{cr})$ as $\sigma_0$ is represented by solid
circles. As is seen these plots are parallel to each other, but
differ by the value about $(1.3\pm 0.3)G_0$.

What is the reason for noticeable difference between the absolute
values of quantum correction, obtained by  different ways? When
evaluating the Drude conductivity we supposed that at $B_{cr}$ the
interference contribution was fully suppressed by magnetic field.
In fact this correction does not equal to zero even at $B_{cr}\gg
B_{tr} $, because at $B\gg B_{tr}$ it decreases with increasing
magnetic field very slowly.\cite{chak,schm,dmit,our1} Therefore,
it is naturally to associate the difference in Fig.~\ref{fig5}(b)
with residual interference contribution to the conductivity at
$B=B_{cr}$. Thus, the proper values of the total quantum
correction are represented in Fig.~\ref{fig5}(b) by open circles,
and the Drude conductivity should be more correctly estimated as
$\sigma_0\simeq \rho_{xx}^{-1}(B_{cr})+(1.3\pm 0.3)G_0$ (see
Table~\ref{tab1}). Note, that the presence of some interference
correction at large magnetic field does not affect the
determination of $(1+3/4\lambda)$ in Section~\ref{sec:hf} because
at $B\gg B_{tr}$ the interference correction is practically
temperature independent.

After we have found the Drude conductivity and the EEI
contribution, we can obtain the interference correction to the
conductivity over entire magnetic field range as
\begin{equation}
\delta\sigma_{int}(B)= \frac{\left(\sigma_{xx}-\Delta
\sigma_{xx}^{ee}\right)^2+\sigma_{xy}^2}{\sigma_{xx}- \Delta\sigma_{xx}^{ee}}%
-\sigma_{0}.
\end{equation}

In Figure \ref{fig6} the magnetic field dependence of $\delta\sigma_{int}$
is presented together with theoretical dependences. One can see that $%
\delta\sigma_{int}(B)$ calculated from Eq.~(\ref{eq07}) as
\[
\delta\sigma_{int}(B)=\Delta(1/\rho^{int}_{xx}(B))
-\Delta(1/\rho^{int}_{xx}(\infty))
\]
well describes the magnetoresistance at low magnetic field (see also Fig.~%
\ref{fig31}), but significantly deviates at $B>0.1$ T. It is not
surprising because Eq.~(\ref{eq07}) was obtained within the
diffusion approximation which is valid at $B<B_{tr}$
($B_{tr}=0.21$ T for this structure). The dependences
$\delta\sigma_{int}(B)$ obtained beyond the diffusion
approximation for back-scattering processes and those taking into
account non-back-scattering processes\cite{dmit} are presented
also. One can see that the experimental data lie closely to the
curves obtained beyond the diffusion approximation. However, our
results do not allow to judge the role of non-back-scattering
processes.
\begin{figure}[tbp]
\epsfclipon
 \epsfxsize=8cm
 \epsfbox{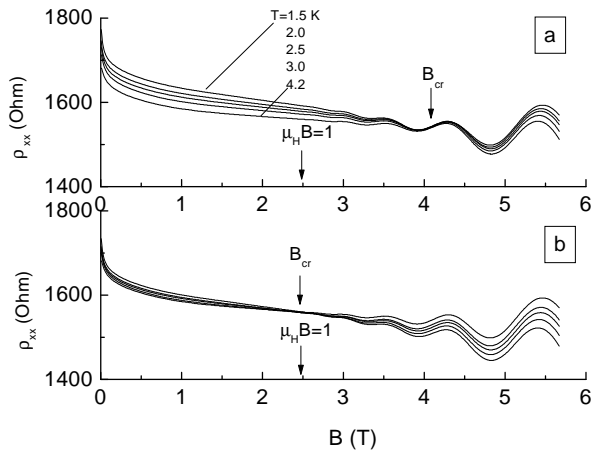}
\caption{Magnetic field dependences of $\protect\rho_{xx}$ at
different temperatures as they have been measured (a), and those
after extraction of the temperature dependence of mobility (b),
structure 4.} \label{fig7}
\end{figure}

\section{Discussion}

\label{sec:disc} As shown in previous sections, all the
temperature and magnetic field dependences for structure 2 are
consistently described by the theory of quantum corrections.
Namely: (i) for $B\gg B_{tr}$, the temperature dependence of
$\sigma_{xx}$ is logarithmic, whereas the temperature dependence
of $\sigma_{xy}$ is negligible; (ii) the low-field
magnetoresistance is well described by the weak-localization
theory with the value and temperature dependence of $\tau_\varphi$
close to the theoretical ones; (iii) the temperature dependence of
the conductivity at $B=0$ is logarithmic and quantitatively
described by the interference and EEI contributions determined
experimentally from the analysis of low and high magnetic field
magnetoresistance, respectively.

Above we have analyzed in detail the results obtained for
structure 2. Similar accordance with the theoretical predictions
has been observed for structures 1 and 3. It allows to determine
the EEI contribution to the conductivity and the values of
$(1+3/4\lambda)$ for different $2k_F/K$ values (see
Fig.~\ref{fig1}). The results for these three structures are seen
to fall on the curve, which is close in shape to the theoretical
one, but lies somewhat lower.

\begin{figure}[tbp]
\epsfclipon
 \epsfxsize=8cm
 \epsfbox{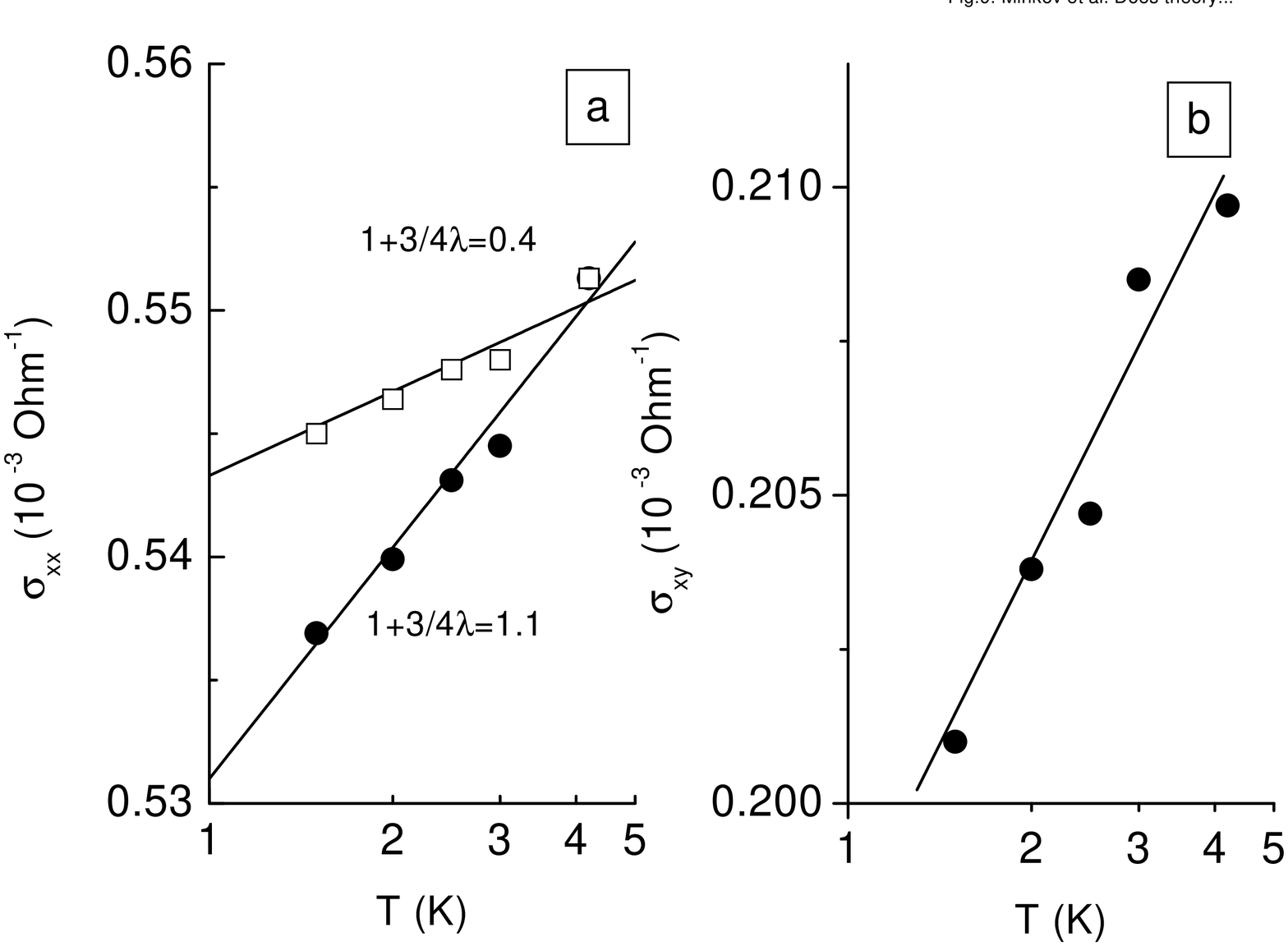}
\caption{$\protect\sigma_{xx}$ (a), $\protect\sigma_{xy}$(b) as a
function of temperature for structure 4, $B=10B_{tr}$. Solid
circles represent the results measured experimentally, open
circles are $\protect\sigma_{xx}$ after extraction of the
temperature dependence of mobility. The lines in (a) are given by
Eq.~(\ref{eq0}), the line in (b) is the guide for an eye.}
\label{fig9}
\end{figure}

What is the possible reason for large scatter in the results
obtained in other papers (see Fig.~\ref{fig1})? To understand this
let us analyze the results for structure 4. It is the structure
with $\delta$-doped barriers, like structure 3, but with higher
electron density (see Table \ref{tab1}). At the first sight the
magnetic field dependences of $\rho_{xx}$ and $ \rho_{xy} $ at
different temperatures (see Fig.~\ref{fig7}(a)) are similar  to
those for structure 2 (Fig.~\ref{fig2}), but unlike structure 2,
the value of $B_{cr}$ is much greater than $1/\mu$. Moreover, the
significant temperature dependence of $\sigma_{xy}$ is evident at
high magnetic field (see Fig.~\ref{fig9}). Such behavior of
$\sigma_{xy}$ is in conflict with the theoretical prediction for
the EEI correction, but yet if one uses the slope of
$\sigma_{xx}$-vs-$\ln T$ dependence at high magnetic field for
evaluation of the EEI contribution, we obtain the value of
$(1+3/4\lambda)$ about $1.1$ that is much greater than that for
other structures. Notice, the temperature dependence of $\sigma$
at $B=0$ for this structure remains logarithmic, but with the
slope about $2.4$ that is essentially greater than the slope for
structures $1-3$: $p+(1+3/4\lambda)=1.35-1.5$ (see Fig.~\ref
{fig5}).

The temperature dependence of $\sigma_{xy}$ at high magnetic field
in structure 4 seems to be incomprehensible. The interference
correction does not depend on temperature at $B\simeq 10B_{tr}$.
The EEI does not affect $ \sigma_{xy}$. Finally, the classical
part $\sigma_{xy}^0$ is temperature independent at such strong
degeneracy of electron gas ($E_F/kT>100$).

It should be noted that in this structure some fraction of the
electrons occupies  the states in $\delta$-doped layers in
contrast to other structures investigated. Analysis of the
magnetic field dependences of $\rho_{xx}$ and $\rho_{xy}$ from the
point of view of two-types of carriers shows that these electrons
do not contribute to the conductivity of the structure. However,
at temperature change the redistribution of the electrons within
the $\delta$-layers can lead to the temperature dependent disorder
and, hence, to the temperature dependence of the mobility of
electrons in the quantum well. Estimations show that as low as 1\%
increase  of the mobility with increasing temperature is enough to
cause the temperature dependence of $ \sigma_{xy}$ observed
experimentally. If we extract this 1\%-changing from $
\sigma_{xx}$ and $\sigma_{xy}$, all the results for this structure
will be in accordance with the theoretical predictions, as for
structures $1-3$: the value of $B_{cr}$ will coincide with $1/\mu$
(Fig.~\ref{fig7}(b)); the value of $(1+3/4\lambda)$ will be equal
to $(0.40\pm 0.05)$; and the slope in temperature dependence of
$\sigma$ at $B=0$ will be equal to ($1.5\pm 0.2)$.

It should be noted that only in Ref.~\onlinecite{r6} the
temperature independence of $\sigma_{xy}$ in high magnetic field
was demonstrated, and, as seen in Figure \ref{fig1} these results
accord well with our data.

Thus, the results for structure 4 show that existence of carriers
in doped layers can bring about additional temperature dependence
of the mobility, and this dependence should be taken into account
when the parameters of the electron-electron interaction are
determined in such type of structures. From our point of view,
disregard of this fact can be one of the reasons for large scatter
of results obtained by other authors (Fig.~\ref{fig1}).

\section{Conclusion}

We have studied the quantum corrections to the conductivity in the
two types of 2D structures: with doped quantum well and with doped
barriers. Successive analysis of experimental data for the
structures where electrons occupy the states in the quantum well
only, shows that all the results are self-consistently described
by the theory of quantum corrections. This allows us to determine
the reliable value of the EEI contribution to the conductivity and
its $k_F$-dependence. It has been shown that the existence of
carriers in doped layers can lead to the temperature dependent
mobility in the structures even at low temperature, when the
electron gas in quantum well is strongly degenerated.

\subsection*{Acknowledgments}

We are very grateful to R.~Kate for her assistance in
collaboration. This work was supported in part by the RFBR
through Grants No.~ 00-02-16215 and No.~01-02-17003, the Program
{\it University of Russia} through Grants No.~990409 and
No.~990425, the Federal Program {\it Physics of Solid-State
Nanostructures}, and the CRDF through Award No.~REC-005.


\begin{thebibliography}
\bibitem{snoska}  From our point of view some distinction of the power $p$
from unity does not demonstrate the failure of Eq.~(\ref{eq08}) for the
temperature dependence of $\tau _{\varphi }$. The fact is that the
experimental values of $\tau _{\varphi }$ are the fitting parameters and
they have to coincide with the phase-breaking time only when the magnetic
field dependence of $\Delta (1/\rho _{xx}(B))$ is described by Eq.~(\ref
{eq07}) exactly.
\end{thebibliography}
\end{document}